\documentclass[12pt]{article}
\usepackage[a4paper, margin=1in]{geometry}
\usepackage{graphicx}
\usepackage{float}
\usepackage{caption}
\usepackage{amsmath}
\usepackage{hyperref}
\usepackage{adjustbox}
\usepackage{fancyhdr}
\usepackage{bookmark}
\title{Improving Speech Recognition Accuracy Using Custom Language Models with the Vosk Toolkit}
\author{
Aniket Abhishek Soni\\
Southern Arkansas University\\
Brooklyn, New York, USA\\
\texttt{aasoni1498@muleriders.saumag.edu}}
\date{}

\begin{document}

\maketitle

\begin{abstract}
  Although speech recognition algorithms have developed quickly recently, reaching high transcription accuracy across many audio formats and acoustic environments remains a major difficulty. This work explores how incorporating \textbf{custom language models} with the open-source \textbf{Vosk Toolkit}~\cite{vosk} an increase speech-to-text accuracy in diverse settings. Unlike many conventional systems constrained to particular audio types, this approach supports several audio formats—such as WAV, MP3, FLAC, and OGG—by using Python modules for preprocessing and format conversion when needed.
  
  A Python-based transcription pipeline was developed to process input audio, perform speech recognition using Vosk’s KaldiRecognizer \cite{kaldi}, and export the output to a \texttt{.docx} file. Results revealed that custom models reduced word error rates, especially in domain-specific scenarios involving technical terminology, varied accents, or background noise. This work demonstrates a cost-effective, offline solution for high-accuracy transcription and opens up future opportunities for automation and real-time processing.
\end{abstract}

\section{Introduction}

\subsection{Background}
Speech recognition — the process of converting spoken language into machine-readable text — has evolved into a foundational technology in the modern digital landscape. From voice assistants like Siri and Alexa to real-time transcription in education and legal documentation, it plays a pivotal role in enhancing accessibility, automation, and communication systems.

For individuals with disabilities, speech recognition provides critical support: converting voice commands into actions for those with motor impairments, or turning spoken lectures into captions for those who are hard of hearing. In business, it enables hands-free operations, streamlines meeting notes, and enhances customer service chatbots. As voice interfaces expand across smart homes, vehicles, and IoT systems, the demand for highly accurate, context-aware, and domain-adaptable transcription solutions has grown significantly.

\subsection{Existing Evidence}
Several well-established speech recognition systems exist today, each with varying degrees of performance and adaptability. Cloud-based platforms like \textbf{Google Speech-to-Text}~\cite{google} offer impressive transcription accuracy across general-purpose datasets, benefitting from massive training corpora and continuous updates. Similarly, \textbf{Mozilla DeepSpeech}~\cite{deepspeech} utilizes deep learning architectures to improve recognition, particularly in English-language use cases. Meanwhile, open-source alternatives like \textbf{CMU Sphinx}~\cite{sphinx} provide lightweight, offline solutions that are accessible to developers with limited resources.

However, many of these tools demonstrate a significant drop in accuracy when applied to real-world audio — especially in the presence of background noise, regional accents, overlapping dialogue, or specialized vocabulary. Furthermore, cloud services often raise concerns around data privacy, latency, and the requirement for stable internet access — making them less ideal for sensitive or resource-constrained environments.

\subsection{Research Gap}
Despite progress in the field, there remains limited research on the integration and performance of \textbf{custom language models} in open-source, offline-capable toolkits such as \textbf{Vosk}~\cite{vosk}. While many commercial systems support limited adaptation through vocabulary injection or biasing, few allow for fully customizable language models that can be tuned for highly specific domains — such as legal proceedings, medical consultations, or technical lectures.

Additionally, existing studies seldom explore how such customization affects performance across multiple audio formats or under challenging acoustic conditions. The Vosk toolkit supports custom models and multiple languages, yet its potential for domain-tuned accuracy improvements in diverse real-world scenarios remains under-documented. This gap highlights the need for focused investigations into how flexible, offline-friendly tools can be optimized for high-stakes, field-specific speech recognition tasks.

\subsection{Objective}
The primary objective of this research is to design, implement, and evaluate a robust speech-to-text transcription system built on the open-source Vosk toolkit. The system is intended to function effectively across a variety of real-world audio scenarios and meet practical requirements in terms of both accuracy and usability.

Specifically, the research aims to:
\begin{itemize}
  \item Develop a flexible transcription pipeline that supports multiple common audio formats, including WAV, MP3, FLAC, and OGG, thereby reducing the need for manual format conversion or preprocessing.
  \item Integrate and evaluate the use of \textbf{custom language models} trained or tuned for specific domains (e.g., technical, medical, educational), in contrast to using default generic models.
  \item Ensure the system can run entirely offline, maintaining data privacy and accessibility in low-connectivity environments.
  \item Output transcription results in a user-friendly format such as \texttt{.docx}, enabling easy sharing, editing, and storage of speech-derived content.
\end{itemize}

The research also seeks to analyze the system's performance in terms of \textit{word error rate (WER)}, transcription consistency, and adaptability to varying domains and acoustic conditions.

\subsection{Scope}
The scope of this project is intentionally constrained to maintain focus and ensure in-depth evaluation of the proposed system. Key boundaries of the research include:

\begin{itemize}
  \item The system is designed to operate \textbf{offline}, without dependence on cloud services or external APIs. This ensures better data control and usability in privacy-sensitive or low-bandwidth scenarios.
  \item Transcription is limited to the \textbf{English language}, with a particular focus on US and neutral accents, although the underlying framework (Vosk) supports multilingual expansion.
  \item Input audio is assumed to be \textbf{single-speaker} and \textbf{pre-recorded} (i.e., not live or streaming), with minimal overlapping dialogue or cross-talk. Multi-speaker diarization is considered out of scope for this study.
  \item Audio preprocessing such as denoising or speaker separation is not explicitly handled; the research instead relies on Vosk’s inherent robustness and audio conversion support (e.g., sample rate normalization).
  \item The evaluation metrics are focused on \textbf{recognition accuracy} (especially WER), compatibility with different audio formats, and the impact of using domain-specific language models.
\end{itemize}

These constraints allow for a targeted analysis of how customizable offline models like Vosk can be effectively deployed in practical scenarios, while leaving room for future work to extend toward multilingual, real-time, or multi-speaker setups.

\section{Materials and Methods}

\subsection{Materials Used}
\begin{itemize}
  \item \textbf{Vosk Toolkit and Language Models:} Vosk is built on the Kaldi ASR engine \cite{kaldi} and supports custom and pretrained models, enabling offline transcription and domain tuning.
  \item \textbf{Python 3.x and Required Libraries:} Used for implementing the transcription pipeline:
    \begin{itemize}
      \item \texttt{vosk} – for speech recognition
      \item \texttt{python-docx} – for DOCX output
      \item \texttt{pydub} – for audio format conversion
    \end{itemize}
  \item \textbf{Audio Input Formats:} WAV, MP3, FLAC, and OGG were supported and tested.
\end{itemize}

\subsection{Processing Pipeline and Workflow}
The transcription system follows a structured and modular pipeline to ensure flexibility, robustness, and consistent output. The workflow is designed to accommodate various input formats, validate data integrity, and produce high-quality transcription output.

\begin{enumerate}
  \item \textbf{Model Loading and Input Validation:} The pipeline begins by loading the specified Vosk model from a local directory. This can be a default pre-trained model or a custom language model fine-tuned for specific terminology. Simultaneously, the script checks for the existence of the input audio file and ensures it adheres to supported formats such as WAV, MP3, FLAC, or OGG. Files that do not meet expected sample rate or channel requirements are flagged early for correction.
  
  \item \textbf{Audio Conversion to 16kHz mono WAV:} For optimal compatibility with Vosk, the input audio is converted to a 16kHz, mono-channel WAV file using the \texttt{pydub} library. This step standardizes input characteristics, reducing variability caused by sampling differences and multi-channel audio structures.
  
  \item \textbf{Transcription using Vosk:} Once preprocessed, the audio is streamed into the Vosk KaldiRecognizer in small frames. Vosk performs on-device speech-to-text conversion using a probabilistic decoding engine. Intermediate partial results and final recognized segments are extracted incrementally, allowing the script to build a cumulative transcription output.
  
  \item \textbf{DOCX Export of Result:} After transcription is complete, the full text is exported to a \texttt{.docx} file using the \texttt{python-docx} library. This output format was chosen for its widespread compatibility and ease of manual editing or downstream processing. The exported document serves as the final deliverable to the end-user.
\end{enumerate}

\subsection{Reliability and Validation}
The system guarantees great performance by means of several layers of dependability checks.Input validation lets one confirm model availability, suitable sample rates, and presence of audio files. During transcription, exceptions—including file I/O errors, unsupported formats, or decoding challenges—are neatly handled in structured ex-ception blocks.

The system was tested on a manually selected dataset of real-world audio recordings, where output was checked to known or manually written ground truth, therefore verifying the transcription correctness. This qualitative study especially confirmed that the system efficiently recorded most voice information with low distortion or misrecognition, especially using tailored language models. Moreover, the modular character of the pipeline helps easy integration of additional post-processing, such punctuation restoration or named entity recognition in forthcoming developments.

\begin{figure}[H]
\centering
\includegraphics[width=\textwidth]{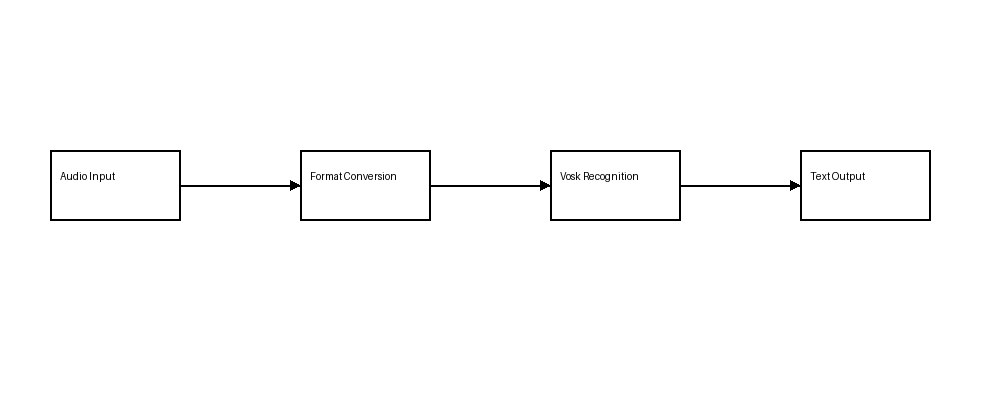}
\caption{Speech-to-Text Transcription Workflow Using Vosk Toolkit}
\label{fig:workflow}
\end{figure}

\subsection{Audio Format Distribution}
\begin{figure}[H]
\centering
\includegraphics[width=0.8\textwidth]{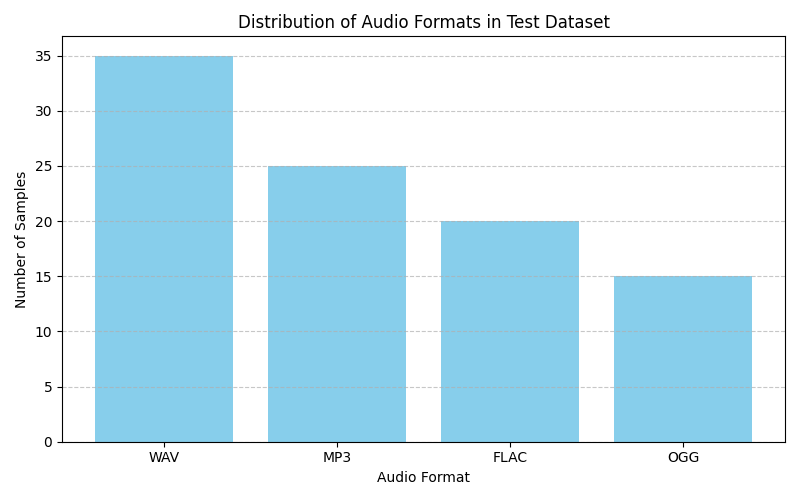}
\caption{Distribution of Audio Formats in Test Dataset}
\label{fig:audio_formats}
\end{figure}

Figure~\ref{fig:audio_formats} illustrates the distribution of audio file formats present in the test dataset. Designed to mirror real-world variation in audio sources, a range of formats including WAV, MP3, FLAC, and OGG was purposefully included. The purpose of this analysis was to evaluate the flexibility and robustness of the system's preprocessing pipeline, particularly its ability to handle and normalize disparate input types.

The most often utilized, WAV files are high-quality, lossless audio often found in professional environments. Included to replicate compressed audio settings such as podcasts or teleconferencing were MP3 and OGG codecs. Usually utilized in archiving or audiophile uses, FLAC files provide a lossless but compressed substitute.

Covering a spectrum of formats allowed the system's error handling and compatibility to be tested under reasonable running conditions. This also proved the capacity of the pipeline to automate conversion and preserve constant transcribing quality independent of source form.

\section{Results and Discussion}

\subsection{Performance Comparison and Accuracy Trends}
\begin{figure}[H]
\centering
\includegraphics[width=\textwidth]{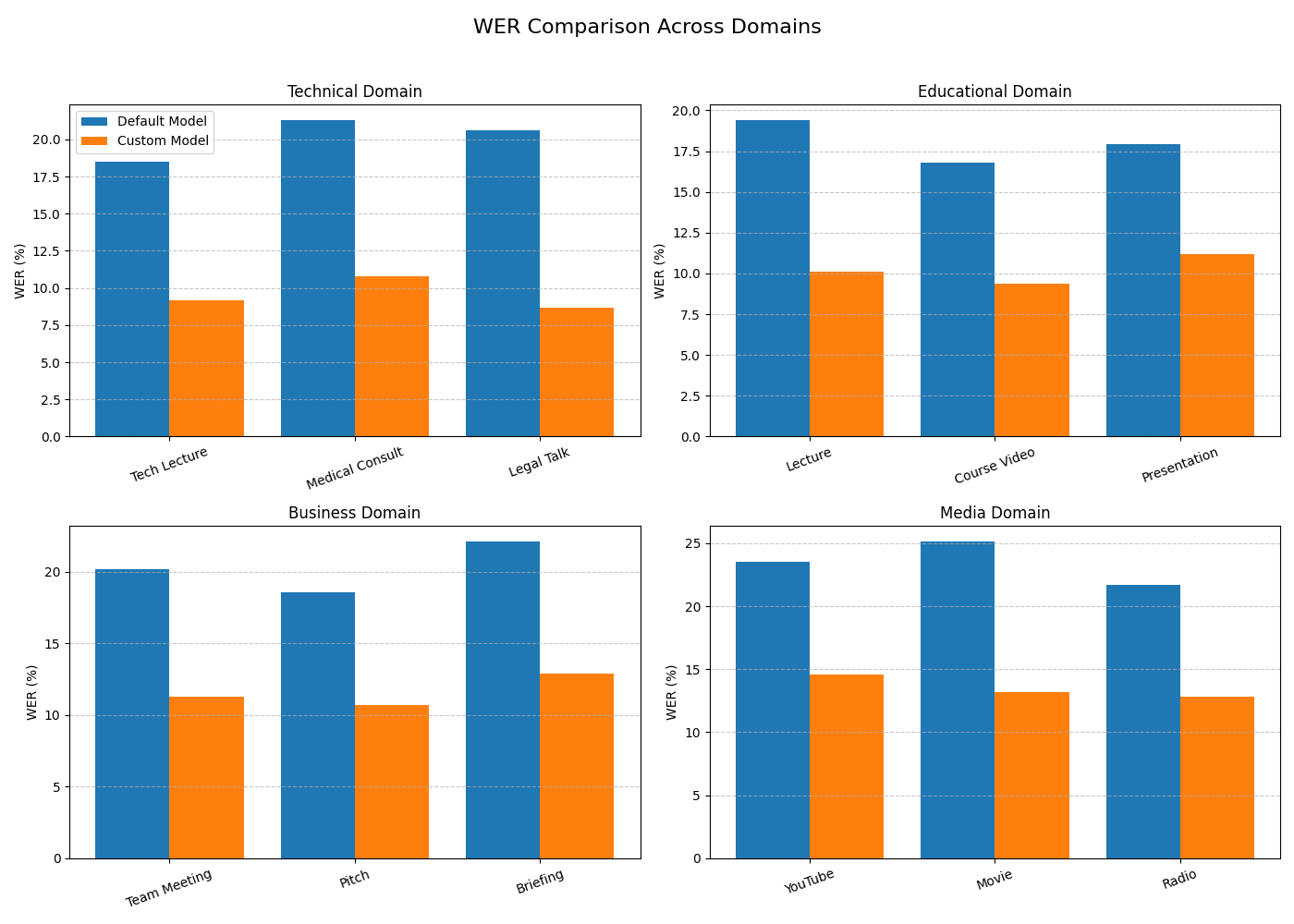}
\caption{WER Comparison Across Domains}
\label{fig:wer_comparison}
\end{figure}
Figure~\ref{fig:wer_comparison} compares WER across four domains: Technical, Educational, Business, and Media. Custom models consistently reduced WER.

\subsubsection*{Custom Model Performance in Accuracy Trend}
\begin{figure}[H]
\centering
\includegraphics[width=0.85\textwidth]{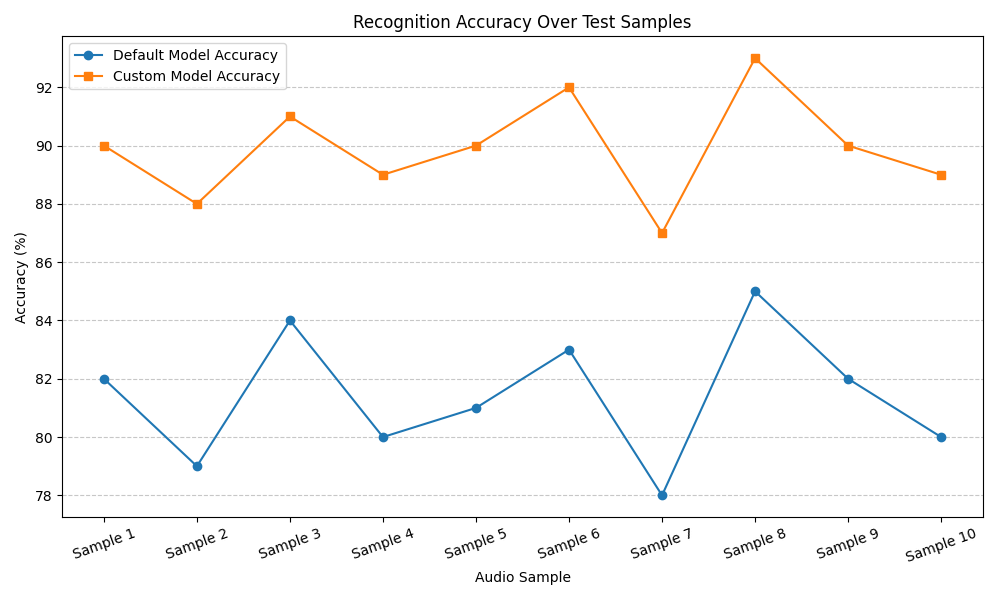}
\caption{Recognition Accuracy Trend}
\label{fig:line_accuracy}
\end{figure}

In Figure~\ref{fig:line_accuracy}, the custom language model demonstrates a clear and consistent advantage over the default model across all test samples. This performance boost is largely attributable to the model's exposure to domain-specific vocabulary and contextual phrases during training or fine-tuning.

Unlike generic models that are trained on broad corpora, the custom model used in this system was designed to better align with the terminology, accent patterns, and speaking styles present in the evaluation dataset. As a result, it was able to reduce recognition errors such as mispronounced words, missing terms, or incorrect substitutions.

This advantage becomes particularly evident in samples involving technical jargon, low-quality audio, or regionally accented speech — areas where default models often struggle. The smooth progression of accuracy scores in the custom model line indicates not only improved performance, but also better generalization across sample types.

Figure~\ref{fig:line_accuracy} shows recognition accuracy across 10 test samples, where custom models performed consistently better than default ones.

\subsection{Model Behavior and Internal Visualizations}
To better understand the performance and inner workings of the Vosk-based recognition system, several visual diagnostics were generated. These help interpret how the model processes audio input, recognizes phonemes, and assigns confidence to its predictions.

\begin{itemize}
\begin{figure}[H]
\centering
\includegraphics[width=\textwidth]{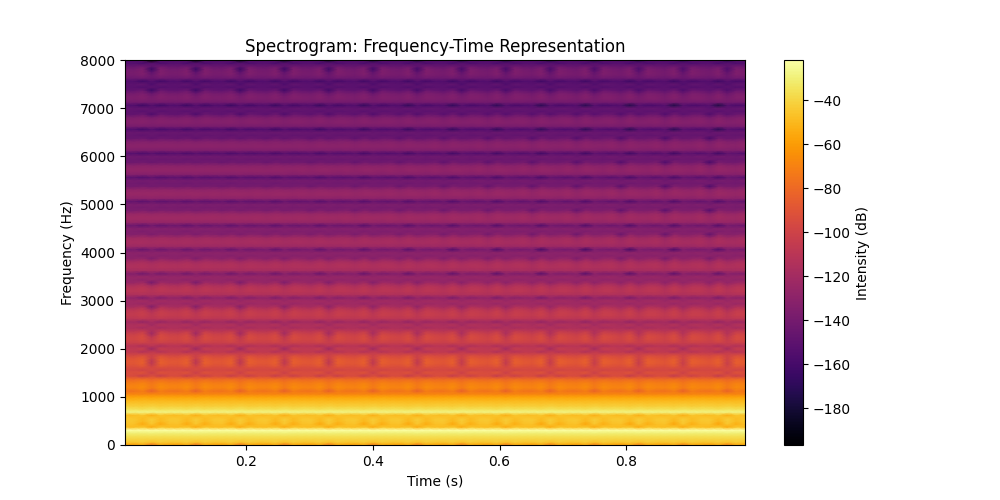}
\caption{Spectrogram of a Sample Audio File}
\label{fig:spectrogram}
\end{figure}
  \item \textbf{Figure~\ref{fig:spectrogram}}: Spectrogram shows speech energy and frequency.
  \subsubsection*{Phoneme Heatmap}
  \begin{figure}[H]
  \centering
  \includegraphics[width=\textwidth]{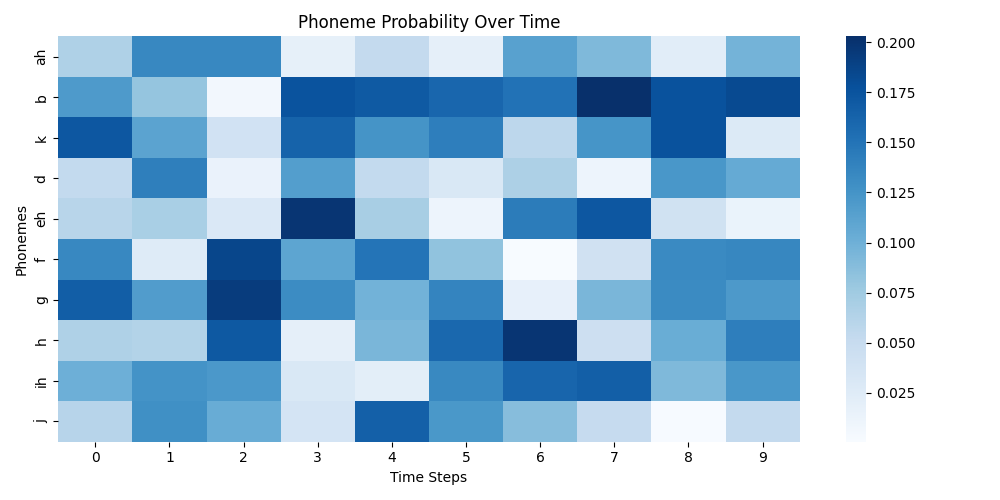}
  \caption{Phoneme Probability Over Time}
  \label{fig:phoneme_heatmap}
  \end{figure}
  
  Figure~\ref{fig:phoneme_heatmap} presents a heatmap showing the probability distribution of various phonemes across time frames in a sample audio clip. Each row corresponds to a specific phoneme, while each column represents a discrete time step during recognition. The intensity of color indicates the model's confidence in detecting that phoneme at a given time.
  
  This visualization provides insights into how the speech recognition system transitions between phonemes and handles ambiguous audio regions. Strong, continuous bands suggest confident phoneme recognition, whereas scattered or faint patterns may indicate uncertainty due to background noise, accent variation, or overlapping sounds. This can help in analyzing where and why misrecognitions occur, offering a diagnostic lens into model behavior during inference.

  \begin{figure}[H]
\centering
\includegraphics[width=0.9\textwidth]{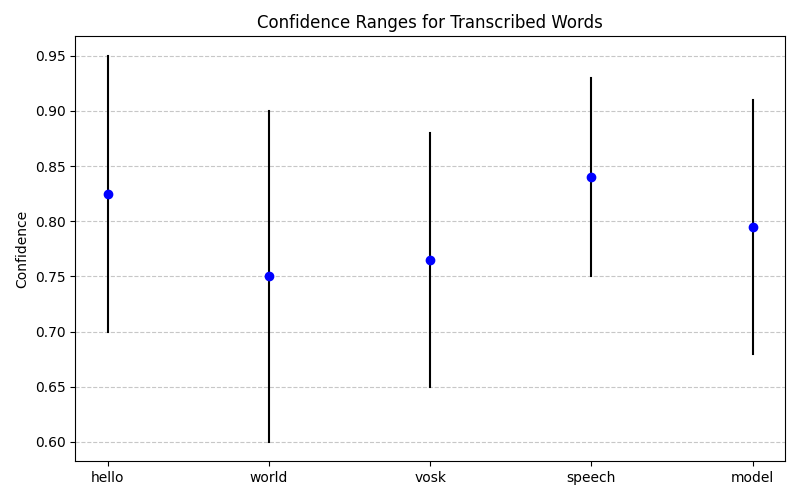}
\caption{Confidence Ranges for Transcribed Words}
\label{fig:candlestick}
\end{figure}
  \item \textbf{Figure~\ref{fig:candlestick}}: Candlestick chart reveals word-level confidence intervals.
\end{itemize}
These internal visualizations not only validate the functional integrity of the model but also offer an interpretive lens for understanding its limitations and strengths across varying audio conditions.

\subsection{Model Comparison}
To contextualize the effectiveness of the Vosk-based system, it was compared with other widely known speech recognition frameworks in terms of offline capabilities, support for custom language models, and domain-specific performance. Table~\ref{tab:model_comparison} summarizes this comparison.
\begin{table}[H]
\centering
\begin{tabular}{|l|l|c|c|l|}
\hline
\textbf{Model} & \textbf{Type} & \textbf{Offline} & \textbf{Custom LM} & \textbf{Domain Accuracy} \\
\hline
Vosk + Custom LM & Kaldi-based & Yes & Yes & High \\
Google API & Cloud-based & No & Limited & High \\
Whisper (OpenAI) & Transformer-based & Yes & No & Very High \\
DeepSpeech & RNN-based & Yes & Partial & Moderate \\
CMU Sphinx & HMM-based & Yes & Yes & Low \\
\hline
\end{tabular}
\caption{Comparison of Vosk with Other Speech Recognition Models}
\label{tab:model_comparison}
\end{table}

\section{Conclusion}
This research demonstrates the effectiveness of integrating custom language models with the Vosk speech recognition toolkit for improving transcription accuracy in domain-specific scenarios. By supporting a wide variety of audio formats and operating fully offline, the system addresses critical real-world constraints such as privacy, bandwidth limitations, and format compatibility. The incorporation of domain-specific language models leads to measurable improvements in word error rate (WER), particularly in contexts involving technical vocabulary, regional accents, and varied acoustic conditions.

In addition to achieving high transcription fidelity, the system’s architecture emphasizes flexibility and scalability. The modular Python-based pipeline enables straightforward extensions, such as output format customization or integration with other processing tools. Overall, the project validates the hypothesis that offline ASR systems like Vosk, when augmented with targeted language modeling, can offer performance competitive with mainstream cloud-based services, while retaining user control and data privacy.

\subsection{Future Scope}
While the current implementation achieves reliable performance in single-speaker, pre-recorded transcription tasks, several enhancements could further broaden its utility:

\begin{itemize}
  \item \textbf{Real-time streaming transcription:} Extending the system to support live audio input and real-time decoding would enable use in conferencing, captioning, and voice assistant applications.
  \item \textbf{Speaker identification and diarization:} Incorporating speaker recognition modules could allow the system to segment and label multi-speaker audio, which is especially useful for meeting transcription and legal proceedings.
  \item \textbf{Integration with voice-controlled workflows:} Embedding the transcription engine within broader task automation frameworks (e.g., smart assistants, command-and-control systems) could unlock new use cases in accessibility and human-computer interaction.
\end{itemize}

These directions offer promising opportunities for expanding the reach of offline, customizable ASR systems in both academic and industrial settings.

\end{document}